# Full-wave computation of SUb-atmospheric Radio-frequency Engine (SURE)

Dingzhou Li[1], Lei Chang[†], Ye Tao[2]

(School of Electrical Engineering, Chongqing University, Chongqing 400044, China)

[†]leichang@cqu.edu.cn

**Abstract**:　Near-space, which covers altitudes from 20 to 100 kilometers, has been receiving more and more attention because of its special strategic value. Airships and high-altitude balloons are two common types of low-speed vehicles that operate in this region. They can be used for jobs like monitoring, communication, and remote sensing, but they need efficient propulsion systems to work well. Earlier, we proposed a new type of electric propulsion system that can ionize the surrounding air to create plasma and produce thrust for near-space vehicles. However, in past experiments, not enough was known about how certain parameters affect power absorption and electromagnetic behavior. Therefore, in this study, we used computer simulations to examine how gas pressure (200–1000 Pa), input power (200–600 W), frequency (13.56–52.24 MHz), and different gas types (Ar, $N_2$, $H_2$, He) influence inductively coupled plasma inside a quartz tube. We especially focused on comparing two antenna designs: one with a single turn and one with five turns. In all the simulations, the single-turn antenna consistently absorbed power better than the five-turns antenna. Higher frequencies significantly influence both plasma power absorption and magnetic field characteristics. The optimal power absorption occurs at a filling gas pressure of 400 Pa. When varying the input power, we observed an initial decrease followed by an increasing trend, which may be related to ionization mechanisms. In comparisons among different gas types, the inelastic collision mechanisms in molecular gases lead to a notable reduction in plasma power absorption efficiency. The results from this work will help guide the design of future experiments for this electric propulsion concept.

**Key words**：　Near space，Plasma thruster，RF discharge，Inductively coupled plasma，Subatmospheric.

## I. INTRODUCTION

Near space, spanning altitudes from 20 (Armstrong line) to 100 km (Karman line) with ambient pressures ranging from approximately 32~5332 Pa, represents a unique transitional zone between conventional aviation and astronautical domains[1]-[3]. This region offers a rarefied atmospheric environment and a geostrategic position that facilitates both upward access to space and downward coverage of the Earth's surface. Vehicles operating in near-space, such as airships and high-altitude balloons, boast significantly longer endurance and broader coverage than conventional aircraft, granting them irreplaceable advantages in applications including persistent wide-area surveillance, communication relay, environmental monitoring, disaster early warning, and scientific exploration[4]-[11]. Consequently, the effective utilization of near-space has demonstrated substantial potential and spurred urgent demand for specialized technologies.

However, the low atmospheric density and low-temperature conditions in near-space pose formidable challenges to propulsion systems. Traditional air-breathing engines suffer from drastically reduced efficiency due to the insufficient oxygen available for combustion. Conversely, conventional electric propulsion systems, such as Hall effect thrusters and ion thrusters, which are designed for high-vacuum space environments, exhibit exceedingly low ionization efficiencies in this sub-atmospheric pressure regime, rendering them ineffective[7]-[8].

To address these challenges, a promising radio-frequency engine concept specifically designed for near-space applications has been proposed[9]. This concept is designated the SUb-atmospheric Radio-frequency Engine (SURE), and a schematic of the SURE device is illustrated in Figure 1. The apparatus primarily consists of a quartz tube, two axially spaced antennas, an RF power supply, and a gas supply system. When RF current is applied to the left-side antenna, it excites an inductively coupled plasma (ICP) discharge beneath it. The right-side antenna is grounded. Due to the coupling between the left-side RF electric field and the plasma potential, a capacitively coupled plasma (CCP) discharge dominates the right-side region[15]. This asymmetric discharge setup creates a potential difference along the tube, thereby accelerating ions to generate thrust. A key advantage of the SURE concept is its ability to utilize the ambient atmosphere as propellant, potentially powered by solar energy via photovoltaic devices, enabling highly efficient ionization without the need to carry traditional propellants.

This mechanism distinguishes SURE from other atmospheric pressure discharge technologies[15] such as corona discharge[16], dielectric barrier discharge[17]-[18], atmospheric pressure plasma jets (APPJ)[19]-[23] cold plasma torch[24], glow discharge[25]-[26], microhollow cathode discharge[27], and surface wave discharge[28]. Many of these alternative technologies involve electrodes in direct contact with the plasma, leading to efficiency limitations caused by sheath formation and electrode erosion[29]. In comparison, SURE does not require magnetic field. Observations indicate that the plasma is primarily generated within the ICP region under the left antenna and propagates bullet-like toward the right. Therefore, this study focuses numerically on the characteristics of the inductively coupled plasma discharge under the left antenna. We used the HELIC to perform calculations and analyze the effects of different antenna turns, gas pressure, input power，frequency，and gas type on plasma power absorption and electromagnetic field characteristics, with the aim of providing a theoretical basis for the experimental design of SURE.

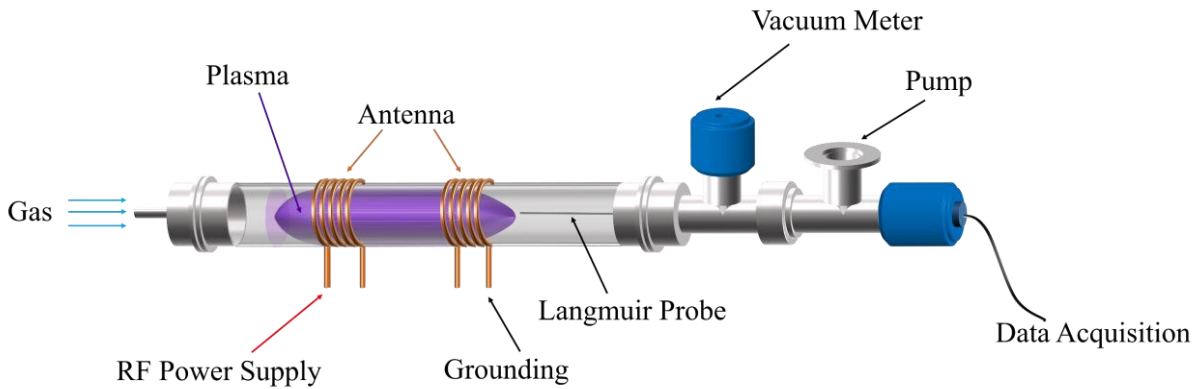

Fig. 1. structure of SUb-atmospheric Radio-frequency Engine (SURE).

## II. HELIC model and calculation program

To investigate the power coupling characteristics between the radio-frequency (RF) antenna and the plasma in the proposed near-space electric propulsion scheme, this study employs the HELIC code for numerical simulations.

Developed by D. Arnush and F. F. Chen[33]-[35], this code is based on Maxwell's equations applicable to radially non-uniform plasmas and the cold plasma dielectric tensor theory. It effectively simulates the excitation and propagation processes of RF waves in plasma. By solving a set of coupled differential equations, it quantitatively analyzes the influence of different antenna turns, gas pressure, input power, frequency, and gas type on plasma power absorption and electromagnetic field distribution, providing theoretical support for experimental design. The foundational theoretical model consists of Maxwell's equations for radially non-uniform plasmas and the standard cold plasma dielectric tensor. These equations can be transformed into the following set of coupled differential equations concerning Fourier-transformed variables:

$$\frac{\partial E_\varphi}{\partial r} = \frac{im}{r}E_r - \frac{E_\varphi}{r} + i\omega B_z, \tag{1}$$

$$\frac{\partial E_z}{\partial r} = ikE_r - i\omega B_\varphi, \tag{2}$$

$$i\frac{\partial B_\varphi}{\partial r} = \frac{m}{r} \cdot \frac{k}{\omega}E_\varphi - \frac{iB_\varphi}{r} + \left(P - \frac{m^2}{k_0^2 r^2}\right)\frac{\omega}{c^2}E_z, \tag{3}$$

$$i\frac{\partial B_z}{\partial r} = -\frac{\omega}{c^2} \cdot iDE_r + \left(k^2 - k_0^2 S\right)\frac{E_\varphi}{\omega} + \frac{m}{r} \cdot \frac{k}{\omega}E_z, \tag{4}$$

The code solves equations (1)-(4) for each set of $k$ values using standard subroutines, with $\omega$ held fixed during each computation. In the cylindrical coordinate system $(r,\varphi,z)$, wave characteristics are described using a first-order perturbation form $\exp[i(m\varphi + kz - \omega t)]$, where $m$ is the azimuthal mode number, $k$ is the axial wavenumber, and $\omega$ is the RF frequency. This set of coupled differential equations is derived from Maxwell's equations, where $k_0 = \omega/c$.

This model handles wave propagation, mode coupling, and power absorption in non-uniform plasma by introducing the plasma dielectric tensor. This enables the wave equation to numerically solve wave excitation, propagation, and absorption. The plasma dielectric tensor is as follows:

$$S = \frac{1}{2}(R+L), \quad D = \frac{1}{2}(R-L), \quad P = 1 - \sum_s \frac{\omega_{ps}^2}{\omega(\omega + i\nu_s)}, \tag{5}$$

$$R = 1 - \sum_s \frac{\omega_{ps}^2}{\omega(\omega + i\nu_s)} \cdot \left[\frac{(\omega + i\nu_s)}{(\omega + i\nu_s) + \omega_{cs}}\right], \tag{6}$$

$$L = 1 - \sum_s \frac{\omega_{ps}^2}{\omega(\omega + i\nu_s)} \cdot \left[\frac{(\omega + i\nu_s)}{(\omega + i\nu_s) - \omega_{cs}}\right], \tag{7}$$

These expressions incorporate parameters such as the plasma frequency $\omega_{ps}$, the cyclotron frequency $\omega_{cs}$, and collision frequencies, reflecting the effects of electron-neutral and electron-ion collision frequencies on the plasma response. The subscript $s$ denotes plasma species, i.e., electrons and ions; the subscript 0 denotes equilibrium values. The parameters $R$ and $L$ represent the right-hand and left-hand circularly polarized modes, respectively. These expressions can accommodate multiple ion species and include displacement current. In this study, a single-charge ion species is considered to simplify the calculations and focus on the influence of antenna structure on coupling efficiency.

The simulation parameters are set to match the experimental conditions of the SURE propulsion device[9], [36]-[37] the plasma region radius is set to 0.025 m, the antenna radius is 0.027 m, and the axial range is $-0.63 < z < 0.63$ m, although it is assumed to be infinite in length. The antenna is positioned axially centered to ensure symmetry. Single-turn and five-turns antenna structures are selected, corresponding to an azimuthal mode number $m = 1$. In this simulation, argon(Ar) was employed as the primary fill gas, supplemented by a comparative study using nitrogen.. All calculation results are normalized to an antenna current of 1 A; therefore, the presented power absorption is a relative result. The background magnetic field strength is 0.001 G, and the wavenumber range is $0 < k < 50$ $m^{-1}$.

## III. Simulation Results

A. Effect of Filling Gas Pressure

To study the distribution of relative power absorption, magnetic field strength, electric field strength, current density, and loaded resistance for single-turn and five-turns antenna coils under different filling gas pressures, relevant calculations were performed using the HELIC code. HELIC is a C++ program used for designing RF plasma sources and guiding experiments; it can simulate the excitation and propagation of RF waves under conditions without an additional magnetic field. In this work, the simulation parameters were set to match the experimental setup. The boundary conditions were configured to finite cavity mode, and the antenna was positioned axially centered to accurately simulate the experiment and ensure a high degree of conformity.

Both the filling gas pressure and the number of antenna turns have a significant influence on the relative power absorption of the plasma. Figure 2 shows the spatial distribution of the relative power absorption under different gas pressures and numbers of antenna turns. As seen in Figure 2, when the input power is fixed at 800 W, the radial power absorption of the plasma first increases and then decreases as the gas pressure increases. The overall radial relative power absorption is highest at a gas pressure of 400 Pa. When the pressure exceeds 600 Pa, the radial relative power absorption drops sharply. The number of antenna turns does not significantly affect the spatial pattern of the plasma's relative power absorption. However, as the number of turns increases, the overall relative power absorption decreases. Similarly, the axial relative power absorption also first increases and then decreases with increasing gas pressure. Overall, the spatial distribution of the axial relative power absorption is consistent with that of the radial relative power absorption. The numerical difference between them arises because the data point for the axial relative power absorption is taken at a radial position of r = 0.02 m. The overall radial relative power absorption increases with larger radius, and the axial relative power absorption is highest at z = 0, which is related to the placement of the antenna. The sharp decrease in plasma relative power absorption observed between 600 Pa and 1000 Pa is related to the discharge behavior under different gas pressures observed in experiments. The discharge likely undergoes a mode transition [38]-[41] within this range—a sharp drop in plasma density and a rise in electron temperature together lead to the reduction in relative power absorption.

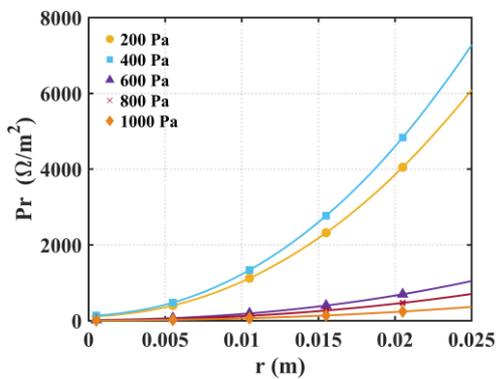 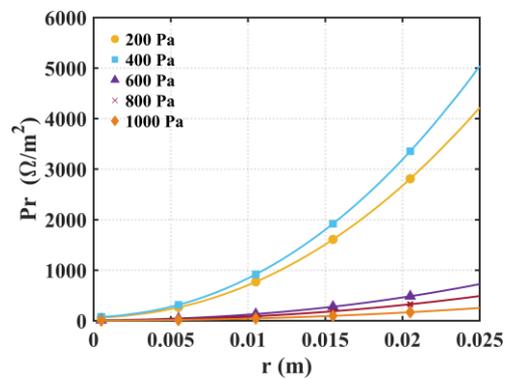

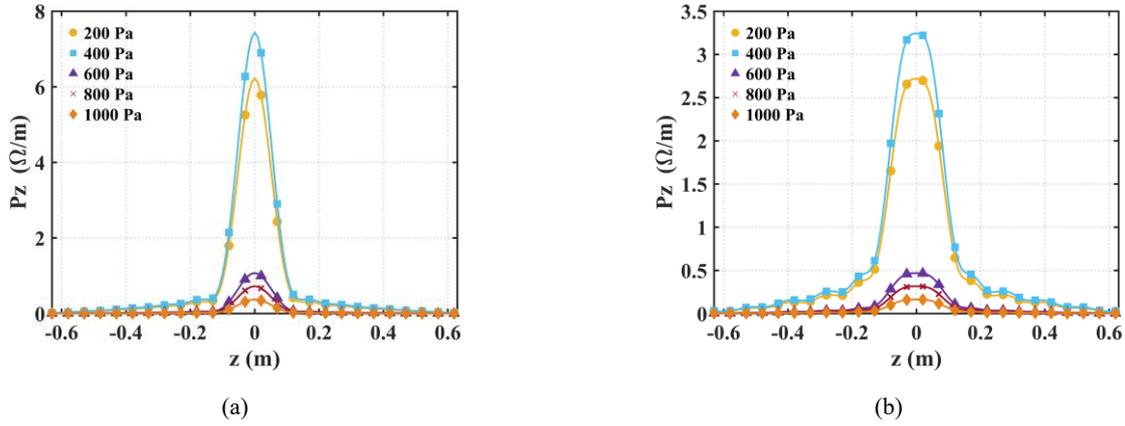

(a)                                                 (b)

Fig. 2. Comparison of Pr and Pz under gas pressure of 200 ~ 1000 Pa：(a) single-turn antenna; (b) five-turns antenna.

Figure 3 shows the distribution of the radial magnetic field strength (Br) and radial electric field strength (Er) along the radial direction for both single-turn and five-turns antenna coils under different gas pressures (200–1000 Pa). As shown in Figure 3, Br initially decreases as the gas pressure increases, but begins to show an increasing trend when the pressure further rises into the higher range. The number of antenna turns has a significant effect on Br. The single-turn antenna consistently produces higher Br values under all pressure conditions, indicating that increasing the number of turns reduces the magnetic field strength generated by the antenna across different pressures. The Er distribution shows a similar pattern. Both single-turn and five-turn antennas give nearly identical simulation results across different gas pressures. At 200 Pa and 400 Pa, the field strength drops noticeably compared to higher pressure conditions, but this effect is only observed with the single-turn antenna. This is because the region under the antenna is dominated by ICP discharge, where Br and Er are primarily determined by the external RF antenna and power supply parameters, such as the coil structure, number of turns, current, and frequency of the RF source. The magnetic field produced by a single-turn antenna is more spread out compared to that of a five-turns antenna. Furthermore, at 200 Pa and 400 Pa, the plasma density is much higher than at pressures above 600 Pa, so the overall magnetic field strength is influenced by the internal plasma under these conditions.

Figure 4 shows the distribution of the radial current density (Jr) excited by antennas with different numbers of turns across a gas pressure range of 200–1000 Pa. It can be observed that, under both single-turn and five-turns antenna configurations, Jr initially increases and then decreases as the filling gas pressure rises. This trend is consistent with the variation in plasma density described in the earlier parameter settings: at 200 Pa and 400 Pa, the plasma density is higher, which improves electrical conductivity and enhances the strength of the induced current. Under higher pressure conditions, the plasma density decreases, leading to a weaker Jr. Further comparison between the different antenna structures shows that the Jr excited by the single-turn antenna is significantly higher than that of the five-turns antenna across the entire gas pressure range. This indicates that the structure with fewer turns can concentrate more RF energy into the discharge region, which is beneficial for enhancing the strength of the induced current in the plasma.

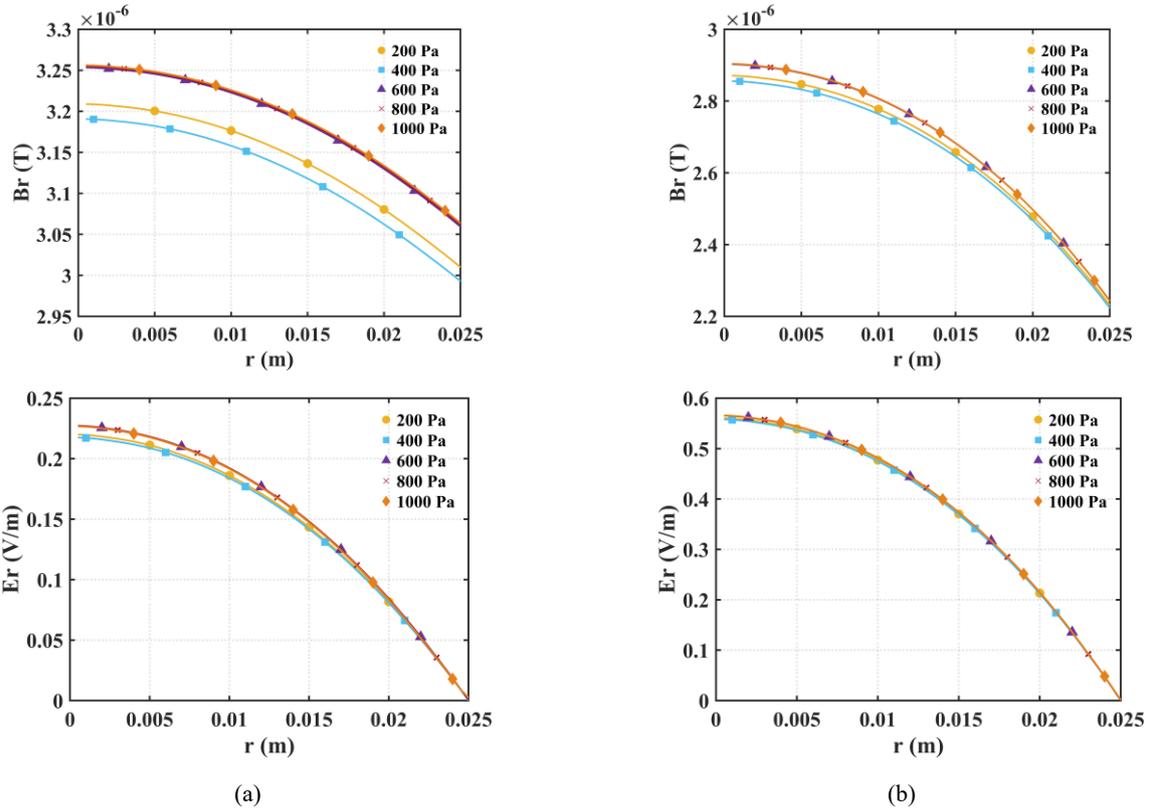

Fig. 3. Comparison of radial Br and Er under gas pressure of 200 ~ 1000 Pa: (a) single-turn antenna; (b) five-turns antenna.

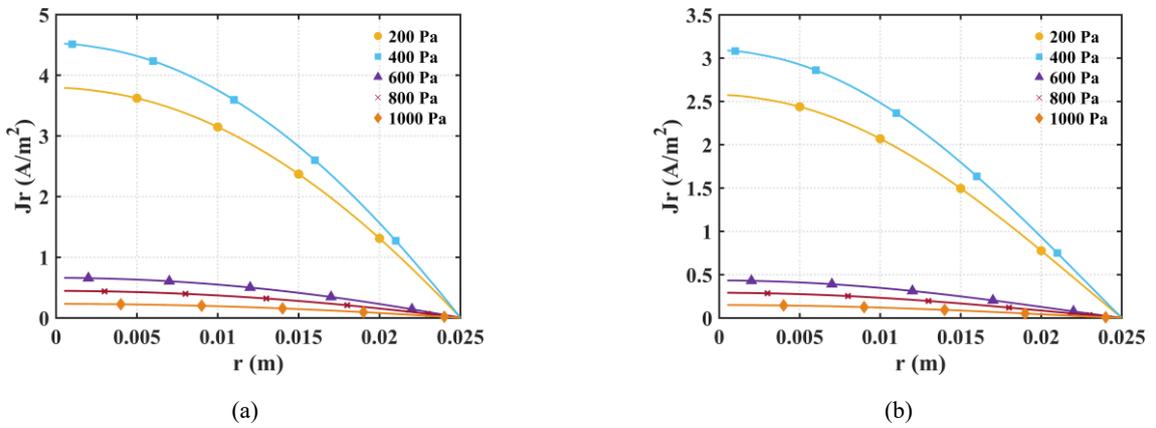

Fig. 4. Radial current density of antennas with different turn numbers under pressure of 200 ~ 1000 Pa: (a) single-turn antenna; (b) five-turns antenna.

The electromagnetic characteristics of the single-turn antenna are higher than those of the five-turns antenna under all pressure conditions. This difference is clearly visible in the wave magnetic field. Figure 5 shows the vertical wave magnetic field for both antenna types at filling gas pressures of 200 Pa, 400 Pa, and 800 Pa. It can be observed that the vertical wave magnetic field strength of the single-turn antenna is significantly stronger than that of the five-turns antenna at the same gas pressure. This observation is consistent with the trends described for previous parameters, confirming the advantage of the low-turn antenna in concentrating energy and exciting magnetic fields, which helps improve overall plasma ionization efficiency. The three gas pressure values—200 Pa, 400 Pa, and 800 Pa—were selected for comparison because they represent distinct conditions under which changes in plasma parameters are most noticeable. For both antenna types, the vertical wave magnetic field strength reaches a relatively highest value at 400 Pa, indicating better energy coupling between the antenna and the plasma at this pressure.

Figure 6 shows the variation of the loaded resistance ($R_nB$) with filling gas pressure, which is used to evaluate the power coupling efficiency between the antenna and the plasma under different operating conditions. As can be seen from Figure 6, for both the single-turn and five-turns antennas, $R_nB$ reaches relatively high values at gas pressures of 200 Pa and 400 Pa, while it decreases significantly in the higher pressure range of 600–1000 Pa. This trend is consistent with the change in plasma density: higher density leads to better coupling, reflected in larger $R_nB$. When the density decreases, the plasma's ability to absorb RF power weakens, resulting in a smaller loaded resistance. It is worth noting that the values of $R_nB$ for the single-turn antenna are consistently higher than those of the five-turns antenna across the entire pressure range. This indicates that the single-turn antenna has an advantage in relative power absorption efficiency. The lower number of turns helps concentrate the magnetic field distribution, enhancing the local RF electric field strength and improving plasma excitation efficiency.

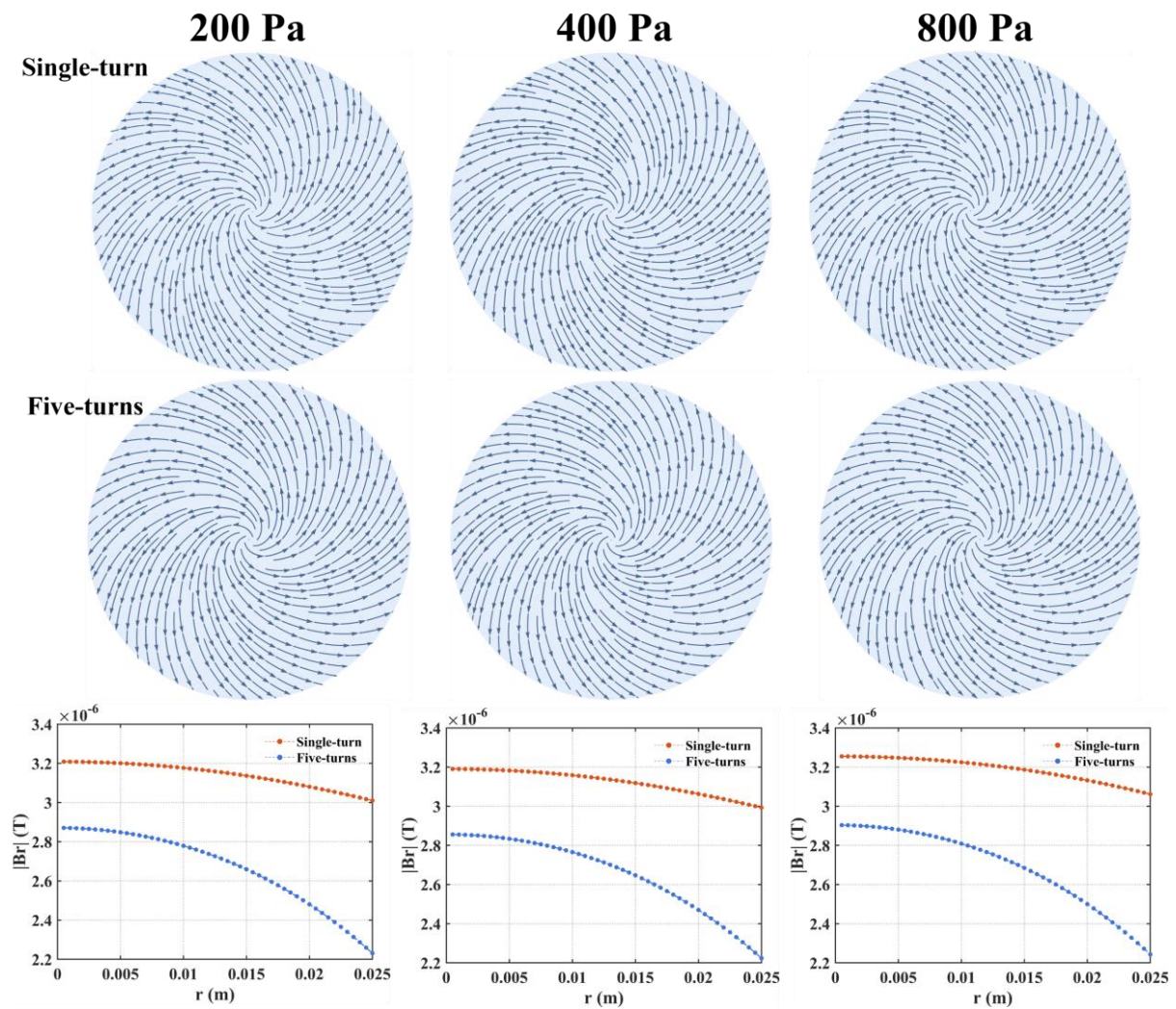

Fig. 5. Comparison of wave magnetic fields between single-turn antenna and five-turns antenna under different filling gas pressures.

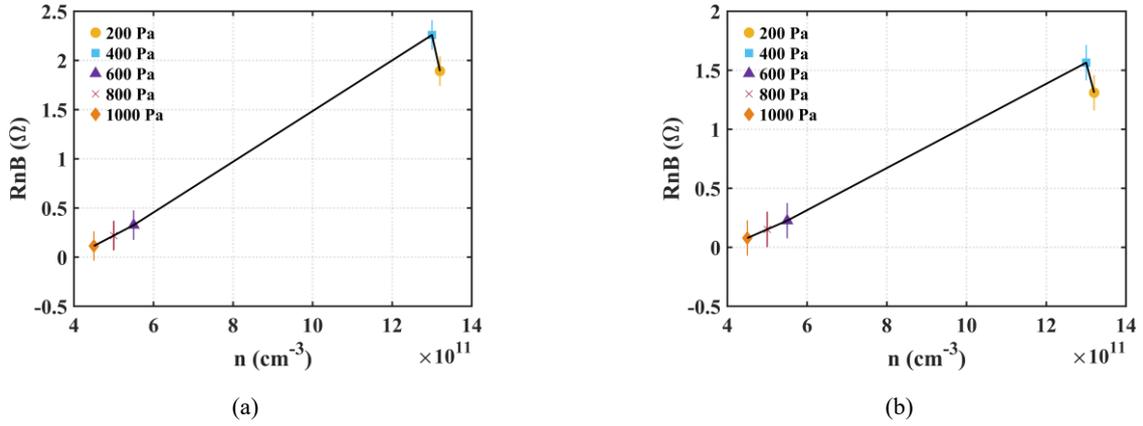

Fig. 6. Comparison of $R$nB under different turn numbers and pressures: (a) single-turn antenna; (b) five-turns antenna.

B. Effect of Input Power

Under different input power conditions, significant changes also occur in the characteristics of the inductively coupled plasma below the RF coil. As concluded earlier from the analysis of gas pressure variations, the spatial distribution of relative power absorption for the single-turn antenna remains consistently higher than that of the five-turns antenna. This observation is further confirmed when the input power is varied.

As shown in Figure 7, for both the single-turn and five-turns antennas, the relative power absorption of the plasma first decreases and then increases with higher input power, showing similar spatial distribution trends. This trend might be related to the ionization mechanism. At an input power of 200 W, the plasma is in a low-power state where single-step ionization dominates. As the input power increases, stepwise ionization begins to prevail. The input power of 400 W likely lies near the transition region from single-step to stepwise ionization, leading to the observed initial decrease followed by an increase in relative power absorption. It is clearly observed that the relative power absorption under the single-turn antenna is noticeably higher than that under the five-turns antenna across all input power levels. This is because the single-turn antenna has lower inductive reactance compared to the five-turns antenna, improving excitation efficiency and thus resulting in higher relative power absorption. Since the simulations were performed with the same background plasma density, electron temperature, and filling gas pressure for each power level, the observed performance difference directly reflects the influence of the antenna structure itself on power coupling efficiency.

A similar distribution trend is also seen in the loaded resistance and current density, as shown in Figures 8 and 9. Both the loaded resistance $R$nB and Jz are higher under the single-turn condition and show a trend of first decreasing and then increasing with rising input power. This indicates that the coupling efficiency between the antenna and the plasma is significantly better with a low-turn antenna and higher input power. It is noteworthy, however, that at an input power of 400 W, the relative power absorption, loaded resistance, and current density for both antenna types reach their lowest values. This further supports the notion that the plasma ionization is in a transition state between single-step and stepwise ionization at this input power level.

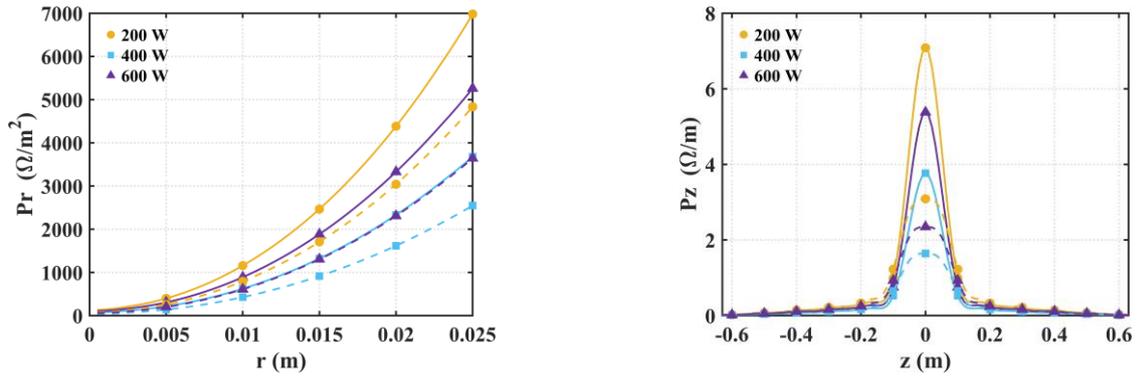

Fig. 7. Comparison of Pr and Pz for antennas with different turn numbers under input power of 200 W ~ 600 W: solid line represents single-turn antenna; dashed line represents five-turns antenna.

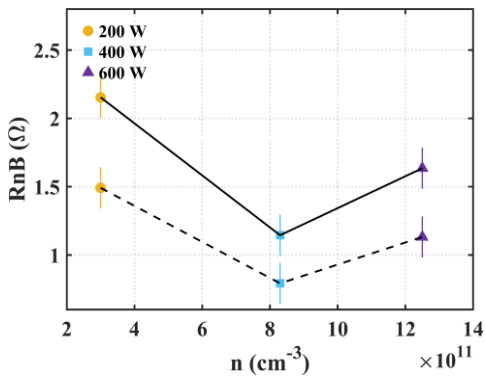

Fig. 8. Comparison of $R$nB for antennas with different turn numbers under input power of 200 W ~ 600 W: solid line represents single-turn antenna; dashed line represents five-turns antenna.

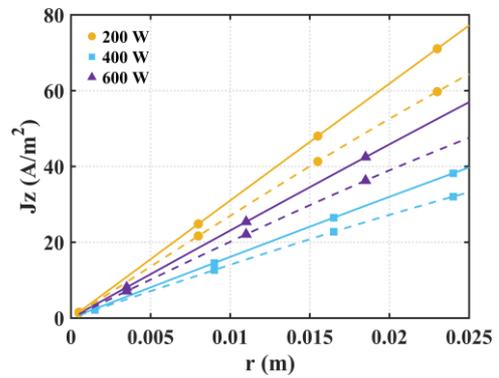

Fig. 9. Comparison of Jz for antennas with different turn numbers under input power of 200 W ~ 600 W: solid line represents single-turn antenna; dashed line represents five-turns antenna.

As seen in Figure 10, when the input power varies from 200 W to 600 W, Br under both antenna types shows a similar spatial distribution: the field strength gradually decreases along the radial direction. Based on earlier simulation results under different gas pressures, the electromagnetic field here is mainly influenced by the antenna structure and RF power supply parameters, which is why the same trend is observed across different power levels. It is worth noting that the decrease in Br along the radial direction is much less pronounced for the single-turn antenna than for the five-turns antenna. This is because the five-turns antenna has a more concentrated magnetic field formed by the superposition of multiple turns, while the single-turn antenna produces a more spread-out field from a single current loop, lacking this superposition effect. In contrast, the multi-turn antenna creates a steeper gradient near the radial edge, resulting in a sharper decrease. According to the law of electromagnetic induction, this attenuation behavior shows an opposite trend in Er.

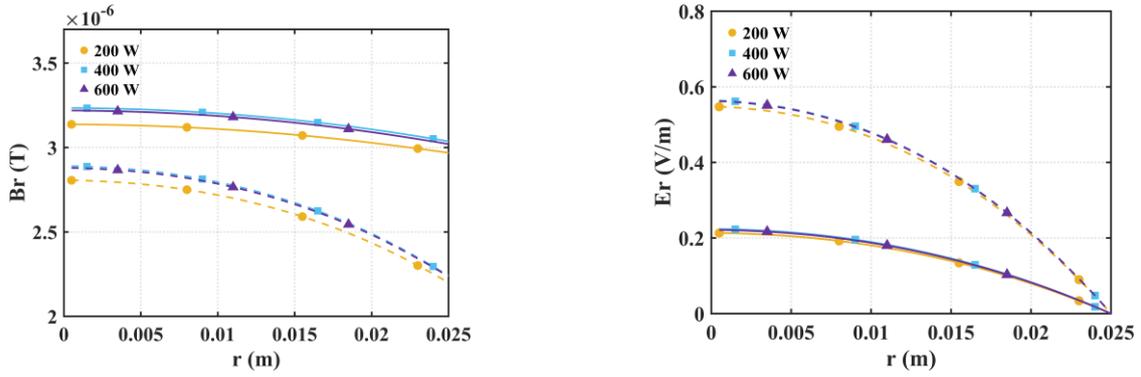

Fig. 10. Comparison of Br and Er under input power of 200 ~ 600 W：solid line represents single-turn antenna; dashed line represents five-turns antenna.

C. Effect of Frequencies

To study the influence of different frequencies on the SURE device, numerical simulations were carried out at four frequencies: 13.56 MHz, 27.12 MHz, 40.68 MHz, and 54.24 MHz . The filling gas pressure was set to 400 Pa and the input power was fixed at 800 W.

As shown in Figure 11, in terms of plasma relative power absorption, both the radial and axial power absorption increase with higher frequency for both the single-turn and five-turns antennas. The increase is nearly proportional to the frequency—the values grow almost by the same multiple as the frequency rises. The radial power absorption increases with larger radius, while the axial power absorption peaks at the center of the antenna (z = 0). This occurs because at higher frequencies, the RF electric field changes more rapidly, allowing electrons to gain more energy from the field. This increases the ionization collision frequency, which improves the plasma's efficiency in absorbing RF power. The growth trend and spatial distribution related to the number of antenna turns are consistent with previous simulations: the single-turn antenna shows higher radial and axial relative power absorption. This further confirms that the number of turns affects the inductive reactance of the antenna, which in turn influences the power absorption.

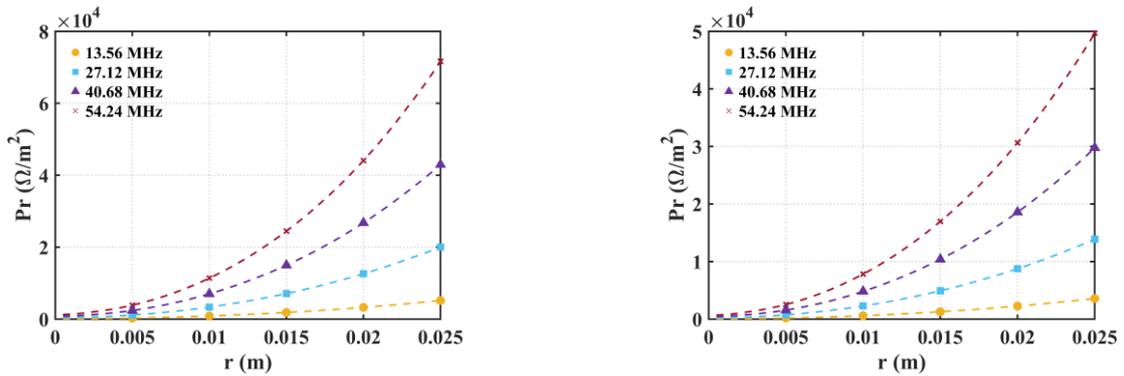

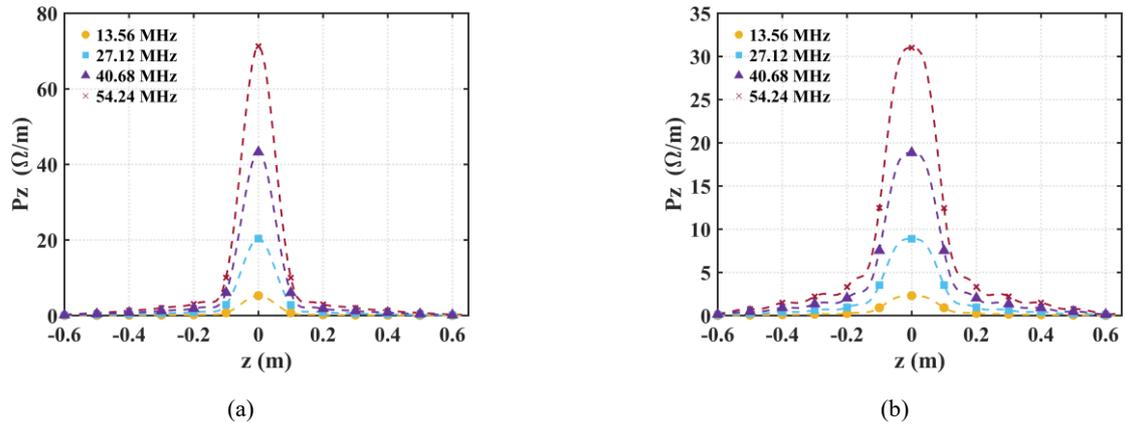

Fig. 11. Comparison of Pr and Pz under different frequencies of 13.56 ~ 54.24 MHz; (a) single-turn antenna; (b) five-turns antenna.

A similar pattern is observed in the electromagnetic field, though the growth and spatial distribution become more uniform and consistent. As seen in Figure 12, Br increases with higher frequency, while Er shows the opposite behavior. Both decrease with increasing radius, which matches the attenuation pattern of the magnetic field generated by the antenna. At the same frequency, the single-turn antenna produces higher Br values, indicating that low-turn antennas have better magnetic field excitation capability at high frequencies. It is worth noting that for the five-turns antenna, the change in Br across different frequencies is relatively small. In contrast, the single-turn antenna shows a more noticeable variation. This may be because, for the antenna with a higher number of turns, the structure has a stronger influence on Br. The magnetic field is more concentrated, so frequency changes have less impact on the field distribution. As shown in Figure 13, the variation and spatial distribution of Jr are consistent with earlier simulations: Jr increases clearly with frequency and decreases with larger radius, but the distribution becomes more uniform. Together with the changes in Br and Er under these conditions, it appears that the variations in the electromagnetic field are more strongly influenced by the antenna structure, especially in the radial direction.

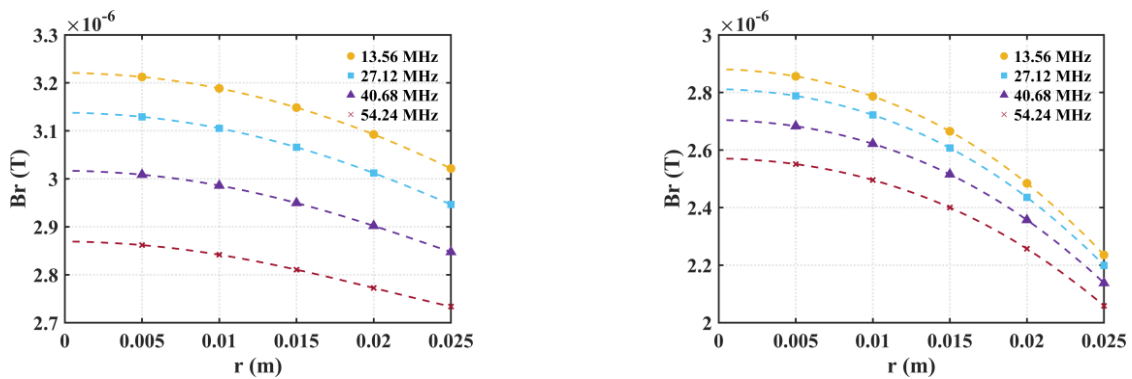

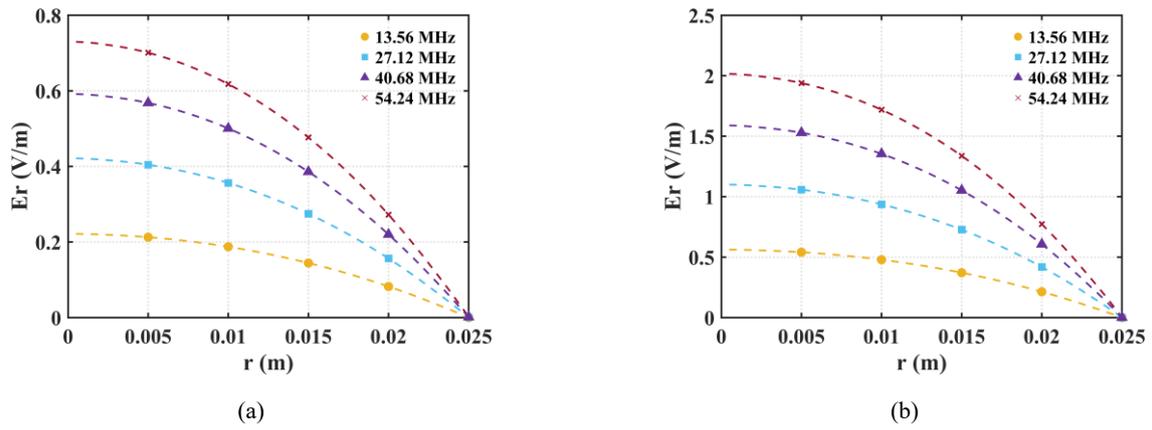

Fig. 12. Comparison of radial Br and Er under different frequencies of 13.56 ~ 54.24 MHz: (a) single-turn antenna; (b) five-turnss antenna.

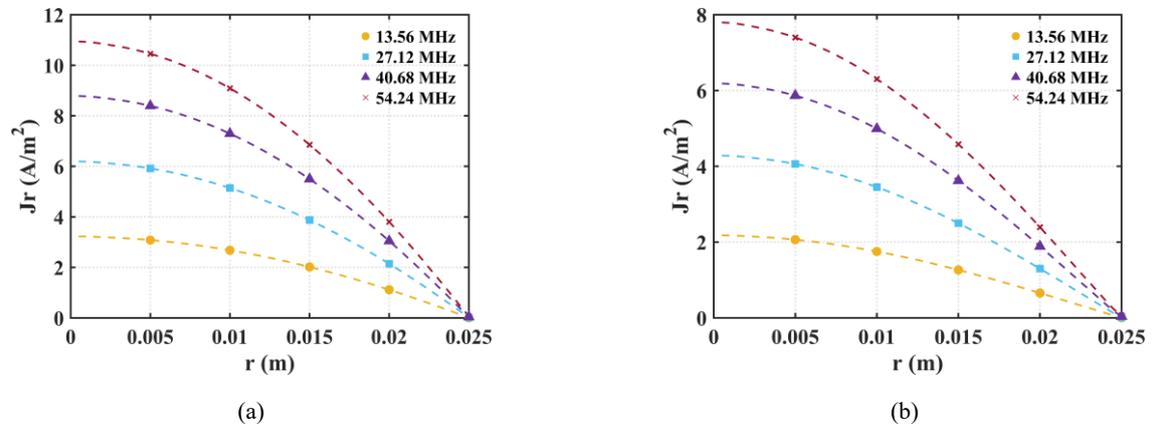

Fig. 13. Comparison of Jr under different frequencies of 13.56 ~ 54.24 MHz: (a) single-turn antenna; (b) five-turnss antenna.

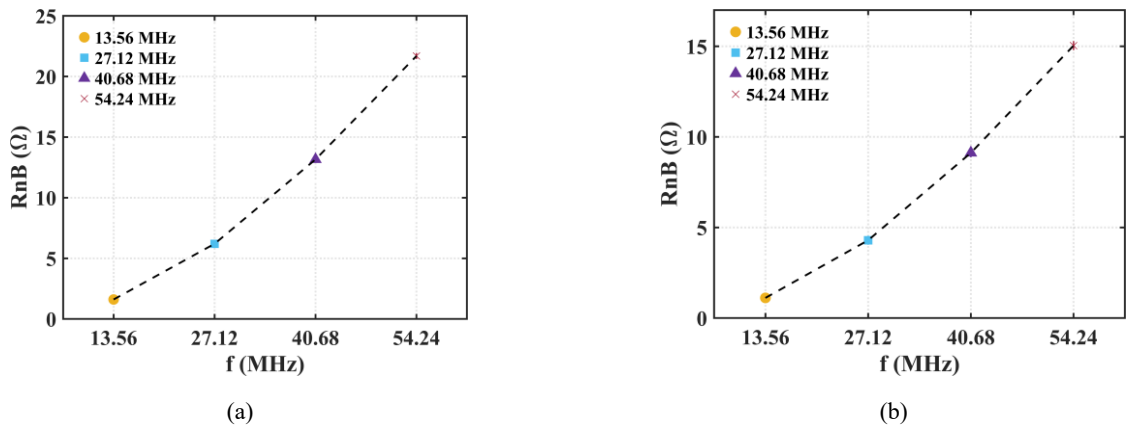

Fig. 14. Comparison of $R$nB under f different frequencies of 13.56 ~ 54.24 MHz: (a) single-turn antenna; (b) five-turns antenna.

From Figure 14, in simulations with both single-turn and five-turns antennas, $R$nB increases significantly with higher frequency. This is because the higher frequency enhances plasma ionization efficiency and wave coupling strength. As before, the single-turn antenna yields larger $R$nB values than the five-turns antenna. This advantage comes from its structural ability to provide a different magnetic field distribution and lower energy loss.

## D. Effect of Gas type

### D.1 Nitrogen-Argon Comparison

In our previous studies, we selected Ar as the filling gas. For future SURE experiments targeting atmospheric air, nitrogen ($N_2$) will be used as the primary simulation gas due to its dominant proportion in air. Other common gases for ICP discharges, including hydrogen ($H_2$) and helium (He), were also simulated for comparison. When switching from one gas to another, the ionization potential of the gas changes, meaning that a higher RF power is generally required to initiate and sustain the plasma discharge. Moreover, differences in the structure and mass of gas atoms or molecules lead to complex collision mechanisms, which also affect the power absorption efficiency of the plasma. As shown in Figures 15 and 16, when the filling gas is switched to $N_2$ and compared with Ar simulation results under different pressures and input powers, it is clear that in the $N_2$ case, the relative power absorption is higher at 200 Pa—opposite to what is observed with Ar — while the input power is kept at 800 W. As a molecular gas, $N_2$ possesses vibrational and rotational degrees of freedom. This introduces additional inelastic collision channels, which become the dominant energy loss mechanism in $N_2$ simulations. In particular, when the filling pressure increases from 200 Pa to 400 Pa, the collision frequency among $N_2$ molecules rises. As a result, more electron energy is lost through inelastic collisions rather than being used for ionization to maintain the plasma. In contrast, Ar is a monatomic inert gas, and its plasma behavior is relatively simpler, involving mainly elastic collisions and ionization. At low pressure, the relative power absorption in Ar plasma depends primarily on electron density and electron-neutral collision frequency. Therefore, as the filling pressure increases to 400 Pa, the relative power absorption rises. When the pressure is fixed at 200 Pa and the input power is varied, the trend in relative power absorption for $N_2$ plasma differs from that of Ar. Due to nitrogen's more complex diatomic structure and the presence of inelastic collision mechanisms, more energy is lost at low input power. In addition, the dominant ionization mechanism shifts with input power, leading to a trend where absorption decreases as power increases. In Figure 17, whether $N_2$ or Ar is used as the filling gas, the spatial distribution of the plasma magnetic field remains nearly the same and is not noticeably affected by changes in pressure. A similar trend is observed when varying the input power, so those results are not shown here. This further confirms that in ICP discharge, the magnetic field strength is mainly determined by the external antenna structure. Changing the type of filling gas has little influence on the magnetic field in the ICP region.

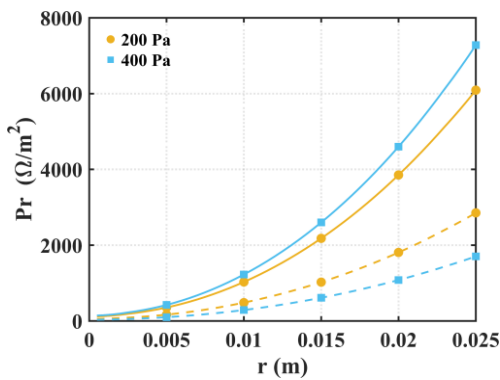

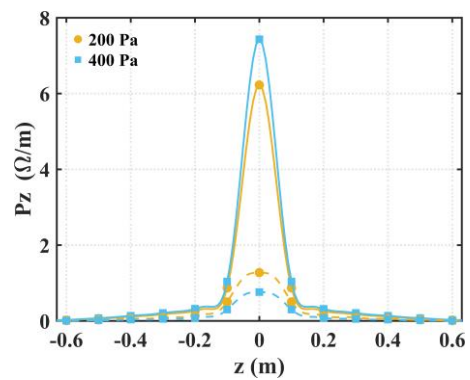

Fig. 15 Comparison of Pr and Pz with a Single-turn antenna under gas pressure of 200 ~ 400 Pa：dashed line represents nitrogen; solid line represents argon.

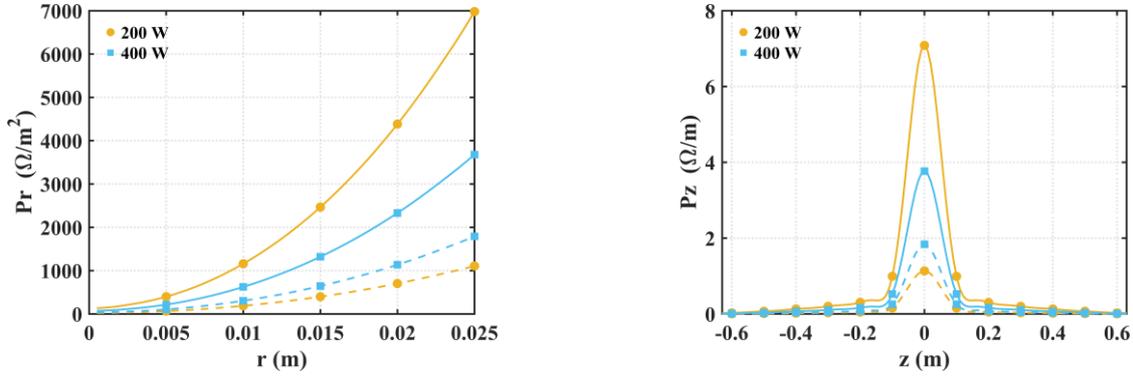

Fig. 16. Comparison of Pr and Pz with a Single-turn antenna under input power of 200 ~ 400 W：dashed line represents nitrogen; solid line represents argon.

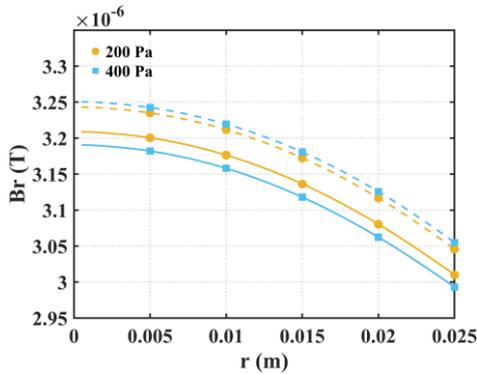 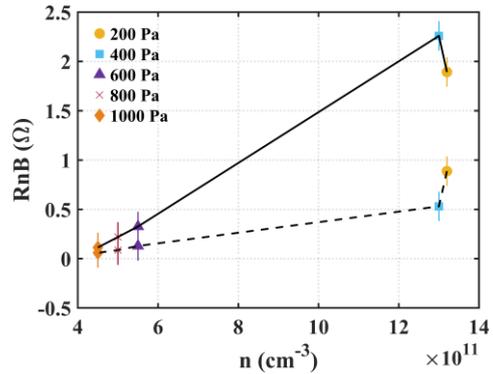

Fig. 17. Comparison of Br with a Single-turn antenna under gas pressure of 200 ~ 400 Pa：dashed line represents nitrogen; solid line represents argon.

Fig. 18. Comparison of $R$nB with a Single-turn antenna under gas pressure of 200 ~ 1000 Pa：dashed line represents nitrogen; solid line represents argon.

D.2 Multi-Gas Performance Analysis

In simulations at different gas pressures, when the working gas is switched to nitrogen, the results are shown in Figure 18. The $R$nB values are significantly higher across all pressure levels. This is primarily due to nitrogen's molecular structure, which leads to lower ionization efficiency, higher collision frequency, and complex dissipative processes, ultimately reducing plasma conductivity. In contrast, argon benefits from its monatomic structure, lower ionization potential, and highly efficient ionization characteristics, which enhance both plasma conductivity and power coupling efficiency.

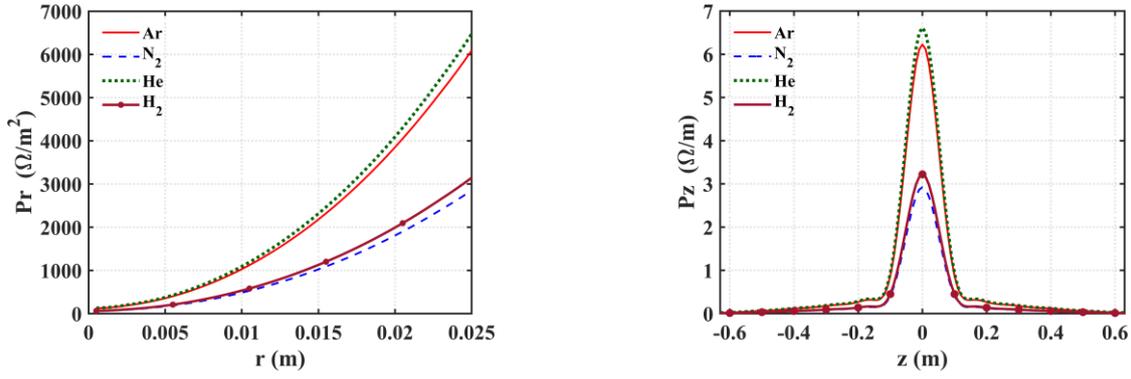

Fig. 19. Comparison of Pr and Pz as a function of gas type (Ar, $N_2$, He, $H_2$) for a Single-turn antenna.

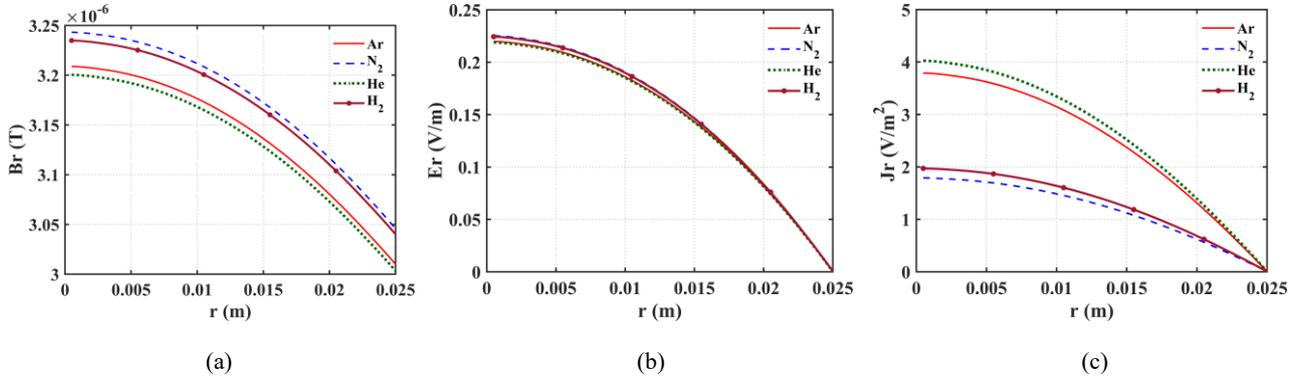

Fig. 20. Comparison of Br, Er, and Jr as a function of gas type (Ar, $N_2$, He, $H_2$) for a Single-turn antenna: (a) Br; (b) Er; (c) Jr.

Here, we introduce additional gas species. Under a filling gas pressure of 200 Pa and an input power of 800 W, the simulation results are shown in Figure 19. The plasma relative power absorption is highest when using He, and the values for Ar and He are very close. In contrast, the relative power absorption for the molecular gases $N_2$ and $H_2$ is the lowest and nearly identical. The atomic or molecular structure of the gas and the electron-neutral collision characteristics influence the plasma's relative power absorption. As monatomic inert gases, He and Ar have similar electron collision cross-sections and energy loss mechanisms. They absorb power primarily through elastic collisions and ionization processes, with almost no additional channels for vibrational, rotational, or dissociative losses. This allows them to convert input power into energy for sustaining the plasma more efficiently, resulting in higher and very similar relative power absorption. Although helium has a smaller collision cross-section than argon, its lighter atomic mass leads to higher electron velocities, resulting in an overall power absorption efficiency similar to that of argon. In comparison, $H_2$ and $N_2$, being molecular gases, possess multiple atomic degrees of freedom. Inelastic collision processes such as vibrational excitation, rotational excitation, and molecular dissociation consume a portion of the input power without directly contributing to ionization or plasma heating. This leads to lower effective power absorption. Especially at low pressure, these additional energy loss pathways in molecular gases cause their relative power absorption to be significantly lower than that of monatomic gases.

In ICP discharge, the magnetic field strength is primarily determined by the external antenna structure, but internal influencing factors cannot be entirely ignored. As shown in Figure 20(a), the spatial distribution of Br remains almost identical across different gases. However, a subtle difference exists between atomic and molecular gases, leading to a slightly smaller skin depth. This enhances magnetic shielding, resulting in a lower internal magnetic field strength. The electric field strength Er is nearly consistent across all gases, as seen in Figure 20(b), indicating that the choice of gas mainly affects the distribution of power deposition rather than the overall electric field. As shown in Figure 20(c), the differences in Jr distribution among the gases stem from the gas-dependent plasma conductivity and skin effect, where plasma conductivity is determined by the effective collision frequency. Being monatomic gases, He and Ar have a lower effective collision frequency. Consequently, their plasma conductivity is higher, and the skin depth is smaller, leading to more concentrated current and higher overall current density with similar distributions. Conversely, the molecular nature of $H_2$ and $N_2$ causes a higher effective collision frequency, resulting in lower plasma conductivity, a larger skin depth, and consequently similar but lower Jr distributions.

## V. Conclusion

The unique environmental conditions of the near-space region impose stringent requirements on propulsion systems. Conventional aerial and astronautical propulsion methods exhibit significant limitations within this domain. Consequently, the development of highly efficient propulsion systems specifically suited for near-space operations has become a critical and urgent challenge. Addressing this need, this study focused on a novel radio-frequency engine, SURE, employing numerical simulations to investigate its plasma power absorption and electromagnetic characteristics. The aim was to provide theoretical support for the experimental design of this propulsion concept.

A simulation model consistent with the preliminary configuration of the SURE device was developed using the HELIC code to investigate the electromagnetic characteristics and power absorption of plasmas for both single-turn and five-turns antenna configurations under varying conditions of fill gas pressure and input power. By comparing key parameters such as relative power absorption, electromagnetic field strength, current density, and loaded resistance, the influence of antenna structure, fill gas pressure, and input power on the inductively coupled plasma was revealed. The simulation results demonstrate that the number of antenna turns significantly affects plasma characteristics. This is attributed to the single-turn antenna's lower inductive reactance, which enables a larger driving current under the same input voltage, thereby generating stronger induced electromagnetic fields and improving energy coupling efficiency. It was also found that the energy coupling effect of the plasma is more effective at lower fill gas pressures. When varying the input power, energy coupling generally increases with higher power but saturates beyond a certain level. Besides argon used as the filling gas, additional simulation studies with nitrogen, hydrogen, and helium as the filling gases were conducted, and comparisons were made.

This study elucidates the influence patterns of key parameters on the performance of the inductively coupled plasma within the SURE model. The findings provide crucial insights and serve as a valuable reference for the subsequent optimization of the SURE experimental design.

## ACKNOWLEDGMENTS

This research is supported by National Natural Science Foundation of China (92271113, 12411540222, 12481540165), the Fundamental Research Funds for Central Universities (2022CDJQY-003), the Chongqing

Entrepreneurship and Innovation Support Programme for Overseas Returnees (CX2022004), and the Natural Science Foundation Project of Chongqing (CSTB2025NSCQ-GPX0725).